# DØ Online Monitoring and Automatic DAQ Recovery

A. Haas, D. Leichtman, G. Watts
*Department of Physics, University of Washington, Seattle, WA 98195, USA*

D. Chapin, M. Clements, S. Mattingly
*Department of Physics, Brown University, Providence, RI 02912 USA*

R. Hauser
*Department of Physics, Michigan State University, East Lansing, MI 48824 USA*

P. Padley
*TW Bonner Nuclear Lab, Rice University, Houston, TX 77251 USA*

B. Angstadt, G. Brooijmans, D. Charak, S. Fuess, A. Kulyavtsev, M. Mulders, D. Petravick, R. Rechenmacher, D. Zhang
*FNAL, Batavia, IL 60510, USA*

The DZERO experiment, located at the Fermi National Accelerator Laboratory, has recently started the Run 2 physics program. The detector upgrade included a new Data Acquisition/Level 3 Trigger system. Part of the design for the DAQ/Trigger system was a new monitoring infrastructure. The monitoring was designed to satisfy real-time requirements with 1-second resolution as well as non-real-time data. It was also designed to handle a large number of displays without putting undue load on the sources of monitoring information. The resulting protocol is based on XML, is easily extensible, and has spawned a large number of displays, clients, and other applications. It is also one of the few sources of detector performance available outside the Online System's security wall. A tool, based on this system, which provides for auto-recovery of DAQ errors, has been designed. This talk will include a description of the DZERO DAQ/Online monitor server, based on the ACE framework, the protocol, the auto-recovery tool, and several of the unique displays which include an ORACLE-based archiver and numerous GUIs.

## 1. INTRODUCTION

In March 2001 the Fermilab Tevatron proton-anti-proton collider started RunII with a center-of-mass collision energy of 1.96 TeV. Both the CDF and DØ detectors and their trigger/readout electronics underwent extensive upgrades to take advantage of the increased center-of-mass energy and luminosity. The DØ L3 Trigger/DAQ group designed and implemented an Ethernet based L3 Trigger/DAQ system (L3DAQ) capable of reading out the DØ detector at a rate of 1 kHz [1]. This paper details two projects that grew out of the L3DAQ upgrade: a monitor data server and a DAQ auto recovery tool.

All DAQ/Trigger systems must have close to 100% uptime. A great deal of effort goes into a system design to achieve this, but inevitably problems occur during operation. Many times problems that stop an experiment's DAQ system are external to the DAQ itself – a digitizer card hangs, for example. In order to quickly diagnose these problems a responsive monitoring system containing complete status data is required.

The monitor system must be able to display system data for both experts and non-experts in a timely fashion to allow for quick problem diagnosis as and debugging. The system must be flexible enough to handle the dual tasks of debugging and commissioning as well as production running. It should minimally impact the performance of the system and also be fairly simple to extend with new monitor data as the need arises.

The monitor system described in this paper successfully met these goals. It is easy to use and has been slowly spreading beyond the L3DAQ project. It has spawned a large range of monitor-data related tools, some of which are described in section 2.4. The monitor system is described in section 2. The program structure, communication protocols, performance, and future directions are discussed.

One of the tools spawned from the monitor project is an auto-recovery system called daqAI. This program gathers monitor information from several different detector components and makes a once-per-second decision about the health of the system. A rule-based expert system the monitor data to make the decision and informs the control room via a text-to-speech synthesizer of a problem. In some cases daqAI can also issue an init or reset to fix the problem. For these classes of problems daqAI has dramatically reduced the time to detect and resume from a problem.

The second section of this paper, section 3, describes this tool, including the expert system, the programming model and possible future modifications.

## 2. THE MONITOR SERVER

The L3DAQ system contains over 150 separate software and hardware components. Understanding the health of the system requires monitor data from all of them. In turn, we have a large number of displays, many designed to address a different audience – experts or shift personnel – or particular tasks -- flagging a rare error condition.

The monitor system's initial design was based on the following requirements. Many of these were based on our RunI experience:

- The addition of new data types to the system must be easy.
- Arbitrarily complex data types must be permitted.
- Allow many copies of the same source object (i.e., 67 readout-crates, 82 nodes, etc.).
- Allow a large number of displays all querying for data.

- Do not make excessive requests for the same data from the same monitor source in short periods of time. In particular if several copies of the same display are running, they should share similar data.

## 2.1. Monitor System Design

We choose to base our system around a Monitor Server. Figure 1 is a block diagram of the system. *Clients* furnish data, and *Displays* request the data. The Monitor Server (MS) sits between the two. The displays do not make direct connections to the clients. All requests in the system are driven by the displays. If a particular client's monitor data is not requested, then the Monitor Server will never request it from a client.

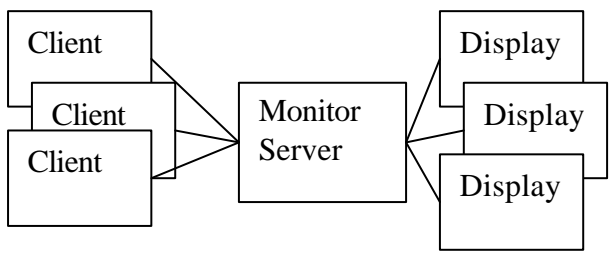

Figure 1 :  A high level diagram of the monitor system. Monitor data flows from left to right, and requests for particular monitor data from right to left. The Monitor Server (MS) caches replies from the clients.

Monitor data is indexed by three keys: the *machine name*, the *monitor type*, and the *item name*. The machine name is the DNS name of the source machine. The monitor type is the class of monitor client – for example a *l3xnode* is a collection of items from a L3 Trigger farm node. Finally, the item name refers to a particular data item. The data returned for an item is arbitrary, and be as large or small as desired (see below). However, the finest grained monitor data request is a monitor item.

The MS stores the most recent reply from each client in a data cache. When a display requests data the MS first checks the cache. If there is a match, the cached data is returned instead making a new request to the client. The display's request may optionally specify a staleness time. If the cache data is older than the staleness time the cache is refreshed with a request to the client. If no staleness time is specified in the request a default time of one second is used.

All communication between monitor system components is over TCP/IP sockets. We use the ACE framework for all sockets programming [2]. This has the added benefit of making the code cross platform (the MS is designed to run on both Windows and Linux).

We have also taken advantage of ACE's multithreaded programming paradigms (see section 2.2). The threading was specifically added to take care of timeouts in the clients with minimal extra programming work. While this will hang a single display's request for data, it will not hang other display's requests.

All connections to the MS are persistent. This avoids overhead involved in setting up new connections. There exists a web-accessible deamon that acts as a go-between to the MS for displays that require a short-lived connection (see section 2.4.2).

The system is designed to recover from crashes and power outages. The protocol requires both the displays and clients initiate their connections to the MS. If the connection is dropped for any reason, the display or client immediately tries to reconnect. It is possible for a TCP/IP connection to be broken without being closed. This occurs most frequently when the MS suffers a power outage. A ping message is sent by the MS to the client every 10 seconds if there has been no data request to the client. If the client doesn't see a ping message every 25 seconds, the connection is dropped. Without this feature all clients would have to be restarted in case of a MS failure.

### 2.1.1. Data Format

All data between the MS and the clients and displays are XML based. The XML structure is shown in Figure 2. At its core, the XML consists of a monitor item name as the XML tag. The reply from the client contains the data as the contents of the tag. We do not maintain a DTD.

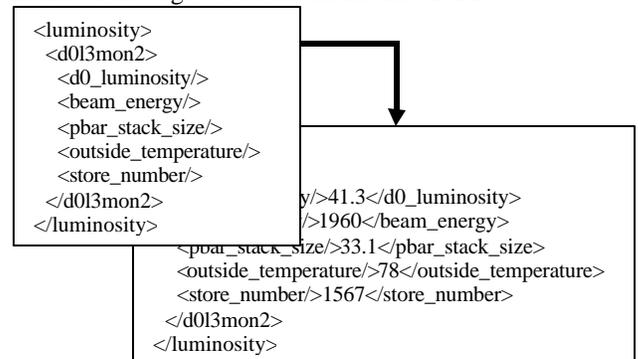

Figure 2 :  Sample client XML request and reply. The upper block contains the XML query sent by the MS to the client, and the lower block represents the reply from the client with the data fields filled in.

The data format has been extremely helpful in debugging and commissioning the system: one can easily read the text that comes back from a monitor item request from a python program or similar.

Each monitor item can contain arbitrary data. The client programmer is encouraged to provide the data in a XML format, but that is certainly not a requirement. While binary data is not legal, it is possible to include almost any arbitrary character using the CDATA XML construct.

All TCP/IP messages are a 32-bit network ordered length word followed by the contents of the XML in ASCII. No binary data is sent in either direction.

## 2.2. Program Design

The block diagram for the MS is shown in Figure 3. The TCP/IP connection handlers are based on ACE's

ACE_Svc_Handler object. This object manages a TCP/IP stream and dispatches data to an input queue as it arrives. There is an individual thread that picks off the data and processes it.

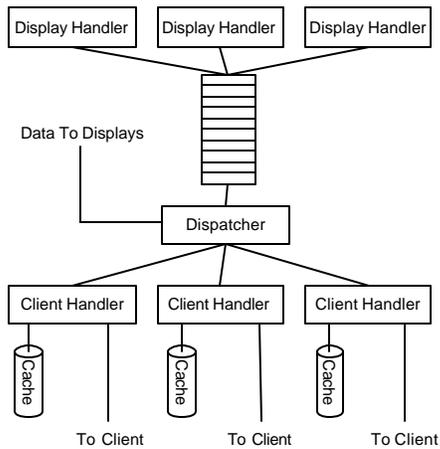

Figure 3 : Block diagram of the MS's object structure. The display handlers feed requests to a processing queue. The dispatcher takes the requests off the queue, parses them, and sends them to the clients for processing. Once all the data has been received back by the dispatcher, the data is sent back to the displays.

The display handlers, dispatcher, and receiver parts of the client handlers all have an associated thread. The display requests are linearly queued. The dispatcher removes them one at a time from the queue and parses the XML. As it parses the display's request, the dispatcher builds a request for each client. Once built, the requests are handed off to the client handlers. After the client handlers have assembled all the requested data and handed it back to the dispatcher, the dispatcher builds the complete reply message and sends it directly back to the waiting display. If the client can't find the data in the cache, it will request it directly from the client. Most requests take less than 150 ms to complete, much less if they involve only cache hits.

All components of the monitor system make connections to the MS. If the MS is not available the client or display will keep attempting to reconnect.

The protocol for the display is very simple. After making the connection to the MS, XML formatted requests are sent, similar to Figure 2. After the MS retrieves the data from its cache or requests it from the clients, it returns a similarly formatted XML document that contains both the data and also all the machines that are of that monitor type. If the data requested is from a machine type with many copies – like a *l3xnode* – then a copy of the data will be returned for each machine. Data from a specific machine can be requested.

The client communication protocol is very similar. The MS will send the client an XML request very similar in format to the display's request. The XML is designed such that the client can just fill in the monitor items one at a time and reply with that information (using the XML Document Object Model (DOM)).

There are numerous timeouts in the system to keep it well behaved even when a client or display misbehaves. If the request cannot be queued by the display handler the display gets an error message. If the request sits on the internal queue longer than one second a timeout message is sent back to the display. A client has 3 seconds to reply to a request for data. If it fails to reply 10 times in a row within 3 seconds, it is disconnected. The dispatcher thread allows 2 seconds for all clients to return their data, and if a client is busy processing the previous request when it starts, it will mark that client as having timed out in the display's reply.

In order to correctly put monitor data in the cache the MS must parse the reply from the client. This is done with a high-speed, zero-copy, hand coded parser.

### 2.3. Client and Display Design

It was recognized early in the monitor system project that simple interfaces would make for wider adoption. The TCP/IP client and display protocols were designed with this in mind: we have also written API's and libraries to implement the protocols. We currently have API's implemented in C++ and python for the client-side protocol, and implementations in C++, python, java, and C# for the display-side protocol.

When a client first connects it advertises its type and machine name by sending an initial XML message. Clients must have a thread listening to the port for incoming messages and must serve them as fast as possible. If the client takes longer than 3 seconds to respond, the MS will flag an error. Repeated failure to respond in time will cause the MS to drop the client's connection.

We have clients in the system that implement the TCP/IP and XML protocol directly. We also have a collection of objects that will take care of all required XML parsing and data conversion. In fact, it is possible to declare an arbitrary instance of a data type to be monitored. Using the common C++ template traits technique the underlying code will render the data to XML whenever a request for the data arrives. Integer counters, for example, can be declared as a template and then used as normal integer in most cases. We have also written a python compiled module that uses a simple name-value pairing to set monitoring variables. The servicing of monitor requests from the MS is invisible to the user. Both API implementations use the xerces XML parser [3].

We have created a similar set of libraries for the display writer. The request to the MS is usually part of the display's main program loop. Various displays often vary what MS items they are requesting depending upon the view the user has chosen to display. The libraries all incorporate XML parsing of one sort or another, though further parsing of complex monitor data item is left entirely up to the display writer. The package most appropriate to the language the user is using is generally used.

A small set of monitor displays are also clients. These frequently collate large amounts of information and publish it back in a collated form. This reduces the amount of data that has to be sent over the wire especially to a display on the other end of a low-end DSL line. The daqAI auto recovery program, described in Section 3, is one such display/client combination.

### 2.3.1. Security

Fermilab is a National Lab, and, as such, all computer systems critical for the operation of the accelerator and the taking of data must be protected by a firewall. The MS is no exception, and thus there is no way to directly contact the MS from outside the firewall. Early on it was recognized that this made the system less useful for remote debugging if displays could not connect. We have received permission to open a single port to a specific machine across the firewall. This second machine receives MS requests and relays them to the MS, and relays the answers back. The relay contains no intelligence, but does do careful buffer length checks, illegal character checks, etc. The relay system is a Windows XP system. All clients must be inside the firewall.

## 2.4. Monitor Displays and Clients

This section contains a brief description of a number of the monitor displays and clients we have running in production.

### 2.4.1. Monitor Clients

The L3DAQ's readout crates contain a Single Board Computer (SBC) that runs the VME readout. The system supplies monitor information on the readout state of each crate, CPU usage statistics, and data transmission failures. The statistics furnished by the SBC to the monitor system requires traversing fairly complex data structures in the program. We have had to use a fast mutex to protect modification by the main SBC program while the monitor data is being collected. Performance of the SBC is not noticeably affected by the locking because the caching feature in the MS reduces the monitor requests to about two per second.

The Level 3 farm nodes are another component for which CPU is a valuable resource. Currently information on incomplete events and CPU usage are generated. There are plans to convert trigger pass statistics and physics performance from another monitor system to the one described in this paper.

The DØ trigger framework (TFW), a non-L3DAQ system, also generates extensive information. This includes all the scalars for the Level 1 and Level 2 triggers and configuration information.

There are also a number of monitor repeaters. For example, we have one system that monitors a web page generated by the accelerator division and scrapes the CDF and DØ luminosity, anti-proton stack size, and even the temperature.

### 2.4.2. Monitor Displays

The principle shift monitor displays for the L3DAQ are written in Java. The designs are based upon the principles outlined in Tuffte's books on the display of graphical information [4]. The main L3DAQ display, uMon, contains a relatively large amount of densely packed information arranged for interpretation by both experts and non-experts. In general we find that though non-expert shifts require about a week to familiarize themselves with the display, they can diagnose a large range of L3DAQ and other subsystem problems with just a glance. Figure 4 shows a portion of the uMon display. A similar display for the L3 CPU farm also exists. The displays were carefully prototyped with simple paint programs and handed around to a small group of experts and non-experts before programming began (PowerPoint, xfig). The displays' designs and usability benefited from this process. This set of displays run on both Linux and Windows.

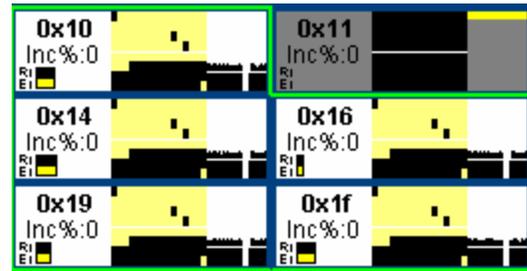

Figure 4 : A small portion of the uMon shifter-monitor display. Each large box represents a single readout crate. The % shows the incomplete event rate for the crate and below it is the status of the L3DAQ route and event queues (on the left, in the white area). The yellow area shows the status of every connection the SBC maintains (there are three farm nodes down). The white area on the right is a rate plot; one small downtime is visible as an inverse white spike.

The L3DAQ also has an expert display based on the freeware version of Qt [5]. This display has a fairly simple main window from which further dialog boxes can be opened. The drill down approach has worked will for getting progressively more detailed information. The display alters its monitor data requests to suit the information it needs to show. Thus it can request detailed, expensive-to-generate information for one or two particular monitor system clients. This display also runs on both Linux and Windows.

We also have written a small Windows systray monitor. This puts a small 32x32 pixel icon in the Windows taskbar that displays the system's health continuously. It has only a rate meter and two green/red circles that indicate general system health. Moving the mouse pointer over the icon will display a small popup with further information. This small display was inspired by Quite Computing principles and has proved useful a useful way for experts to watch L3DAQ while doing other work.

The systray monitor is often run on a portable, which isn't always connected to the internet. It is more convenient to use an http based interface for this monitor tool. There is a web site that acts as a front-end for the monitor server. The web site, called *l3mq*, also allows developers debugging the system to issue monitor queries without having to write code. It is also possible to store a query and reissue it by accessing a single URL. Finally, the web site collects statistics from the MS about which items have been requested and maintains a database. The Web Site user can then add documentation. In the future this will automatically be turned into a manual of all monitor items available in the system.

We also have an archiver program which issues a query once every 15 seconds and writes the results to a large data file. A web interface allows one to make time-based queries and plots online. The format was originally a root file, but indexing proved difficult. The database was converted to Oracle, but the amount of stored data proved to be too large for the system. We are planning to switch to a mixed system: root to store the raw data, and Oracle to store the index information.

## 2.5. Performance

The MS has been in operation for almost 2 years. The monitor system currently runs on a dual 1.2 GHz CPU Linux based system with 0.5 GB of RAM. Typical usage has it querying clients for 0.5 MB/sec and replying to displays with 1-2 MB/sec of data. The CPU is usually 25% busy. The display typically has 150 clients connected and over 70 displays.

The data format is ASCII and not compressed. During our initial running we had a system-wide query that was delivering 1 MB/sec to each copy of a particular display type. We developed an ASCII encoding to compress the data so that less than 100 KB was delivered to each display.

The MS can have more than 100 threads executing at a time. We have not noticed degradation in the CPU or the performance as a function of the number of threads running on the machine.

Our original implementation of the MS used Xerces to parse the replies from the clients. This proved to be a CPU bottleneck. Since the MS doesn't care about the contents of the data item, it didn't make sense to spend CPU cycles parsing it. We wrote our own custom parser that takes advantage of the known format of the XML replies. This reduced the CPU utilization by an order of magnitude.

## 2.6. Future Directions

The MS is a stable product and rarely crashes or has modifications made to its source base. We have altered some of the communication timeouts as the rest of the system grows more stable (lengthening them).

In future we may desire to have more monitor displays or clients. One possibility is to make a hierarchy of caching monitor servers. Each MS queries the MS below for information. This is particularly attractive if you have a large number of a particular client with a fairly stable query.

It is also possible to run with multiple MS, each one devoted to a particularly large sub-system. The implementation for this is a matter of configuring what machines/ports the MS runs on.

## 3. THE DAQAI AUTO RECOVERY UTILITY

There are small classes of DAQ problems in a large experiment like DØ that are easy to recover but cause significant downtime. For example, DØ had a bug in some readout crate code. The programmer was unable to fix the bug immediately because they were stuck outside the country (visa difficulties, post 9/11). The result was 30-120 seconds of downtime every 10 minutes. The length of downtime was a strong function of the wakefulness of the shift personnel. The problem was easily recognizable and also easy to recover from: a single init command needed to be issued.

This experience and several other similar ones were daqAI's genesis. The utility is designed to recognize a number of specific problems, and, if possible, recover the DAQ system to continue taking data without shifter intervention. The daqAI utility also informs the control room via a text-to-speech interface what it is doing and what problem it has found and keeps a shift summary. It is important that the control room be informed of daqAI's actions other wise the shifter and the program could work at odds.

The system is designed around a fact-based embeddable expert system, CLIPS [6]. The system is built around a rule-inference system. Monitoring data is collected from the MS and converted to facts. The facts drive the rule engine, which will in turn execute embedded subroutines and functions or define new facts which will, in turn, cause more rules to execute. Functions are defined that can effect the desired changes. Figure 5 shows a high level architectural diagram of the system.

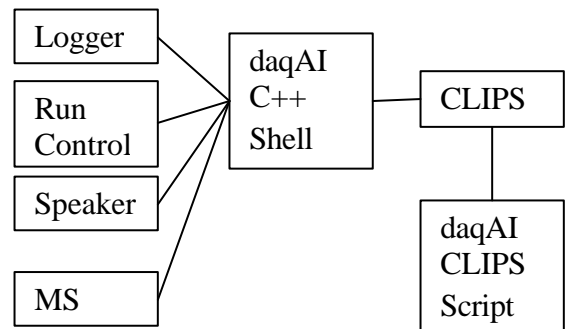

Figure 5 : The daqAI architecture. A C++ shell mediates the actions and inputs of the CLIPS script and all components external to the system. Connections to the Logger, Run Control, the text-to-speech synthesizer, and the monitor server are all by TCP/IP.

daqAI is both a MS client and display. It uses the display features to gather the data it uses to make its decisions, and the client features to publish its internal state, actions, and log.

## 3.1. Program Structure

daqAI, the C++ program, is a shell. Embedded in the shell is the CLIPS system. At runtime a CLIPS rule script is loaded and run thus allowing us to change its behavior without having to rebuild the system. Figure 5 shows the general design.

### 3.1.1. The C++ Shell

daqAI is designed around a loop that executes forever. The loop is repeated approximately once per second. The

```
(defrule b_daq_rate_low "Is rate too low?"
        (daq_rate ?rt&:(< ?rt 50))
        =>
        (assert (b_daq_rate_low ?rt))
)
```

Figure 6: A simple CLIPS rule. This rule will assert a new fact, *b_daq_rate_low* if the fact *(daq_rate x)* is present and x is less than 50.

loop first makes a MS request to gather the monitor data. The system then resets the CLIPS system to its cleared state and defines facts corresponding to the monitor data. The translation algorithm is a fairly simple text based one. The CLIPS inference engine is then run. During the engine execution rules make callbacks into the daqAI shell to request logging output or request an init of the L3DAQ system. The daqAI shell does nothing during the callbacks other than to mark they have occurred. Once the inference engine has run to completion, the daqAI shell examines the DAQ init requests, log requests, text-to-speech requests, etc., for new ones that weren't present on the previous iteration. The new requests are acted upon, the old ones ignored. Any requests that were made last iteration but not the current iteration are so noted, though no external action is taken. Finally, monitor variables and information are generated and published for any MS requests. The loop then repeats.

The CLIPS rule engine is started fresh for each iteration through the main loop. All previous knowledge is erased from the engine at the start of the loop. Thus, the system arrives at the same set of conclusions each time through the loop as long as the inputs remain the same. Of course, if a problem occurs requiring a reset to be issued, one would expect the system to issue a reset each time through the loop. As shown in Figure 5, daqAI C++ is a shell around the CLIPS system. It watches the requests that are made by the CLIPS system and takes action only when it observes a change. So the first time a reset request is made, the shell will actually issue the reset. If the same request is made on the next iteration, no reset will be issued to the DAQ system. This pattern is followed for all actions.

The system could have been designed to remember facts from iteration-to-iteration. A facts based system is well suited to noticing a set of facts in combination and flagging them. However, the code to recognize that this set of conditions no longer exists, but did just before is not clear or easy to write. The problem domain is also well suited to the idea of a fresh start each iteration: this version of the system isn't designed to watch for patterns in the time-domain: just the presence of a set of condition.

There are situations that require some memory from previous iterations. A number of monitor variables tend to give false reading for a very short period of time, for example. daqAI must make sure the monitor variable is out of range for an extended period of time. daqAI contains a number of useful constructs give the script crude model of time. There are timers that will count up as long as a special function is called each iteration. If it isn't called, the timer resets to zero. The timer is available as an input for rules as a fact. There are also counters and even arbitrary facts that can be set and thus *remembered* from iteration to iteration.

The main loops plug-in architecture allows for communication to an arbitrary set of external devices. Currently these include the main DØ Run Control program, log files, a control room text-to-speech synthesizer, the official control room logbook, and email. Each time daqAI identifies an error it will assign it a name. Once a downtime condition has been resolved a complete report is added to the official online logbook, where it can be viewed by any member of DØ. At the end of each shift daqAI reports what errors occurred and their total downtime. This gives an accurate accounting of downtime at DØ.

### 3.1.2. The CLIPS Language

CLIPS rules are stored as a text file. The structure of the script is completely up to the programmer. The language is rich containing not only rule constructs but also objects. The daqAI script makes use only of the rule style. For complete documentation see references [6]; this section contains a very brief introduction to rule based programming in CLIPS.

The runtime environment contains a list of active facts. Each fact has a name and an arbitrary number of arguments. For example *(daq_rate 33)* might indicate the L3DAQ rate is currently running at 33 Hz, and *(bad_muon_roc)* might indicate that a muon readout crate has gone bad. In the daqAI program, the C++ shell defines a list of facts that are a direct translation of the monitor data.

The CLIPS program is a list of rules. Rules have preconditions and actions. A rule *fires* when the preconditions are met. Preconditions are usually pattern matches involving facts. Figure 6 is a simple example. This rule will fire only if the *daq_rate* is less than 50 Hz. In this case, this rule will assert a new fact: *(b_daq_rate_low 33)* given the initial presence of the *(daq_rate 33)* fact.

Very powerful programs can be built out of this simple set of constructs. Facts represent the current environment and rules represent the knowledge.

### 3.1.3. The CLIPS daqAI Script

The daqAI CLIPS script is the heart of daqAI. Its rules contain the hand coded knowledge of the problems it recognizes and the actions it should perform upon their recognition.

The layout of the script is in several tiers, each tier feeds the next. The lowest tier contains the facts that are directly converted to facts by the daqAI C++ shell. The second tier contains very basic inferences from the raw data. For example, it contains a rule testing for a low L3DAQ event rate. The third tier contains problem recognition rules and often involves several second tier inputs. The third tier also insures that the existence of a problem is worrisome. This prevents daqAI from trying to control the system during commissioning, for example. The fourth tier contains the action rules. These issue commands to run control, log messages, send text to the speech synthesizer (and old DECTalk DTC01 machine).

### 3.2. Performance

The current version of daqAI has been in operation for almost a year without major modifications. When the daqAI program first started running the DØ DAQ system had a number of problems that daqAI was able to fix much faster and more consistently than most shifters. The result was data taking efficiency went from about 75% to 85%. Currently, in June of 2003, the DØ DAQ system is much more stable and as a result daqAI's direct impact is less. One of its more important functions now is to create the shift summaries and list the individual downtimes.

The current version of the daqAI CLIPS script uniquely recognizes 8 different problems, in addition to the general *Unknown* downtime. For 4 of them there are established automatic recovery procedures. The *unknown* problem classification indicates a problem that daqAI doesn't explicitly recognize.

One key to a system like daqAI's success is the wealth of monitor information available to it. Adding new data sources to our MS system only increases the potential of a tool like daqAI.

### 3.3. Comments on Usage

Though daqAI has proved quite useful, it is not without its problems. In particular, the amount of work required to identify a common problem and implement an automated fix can be daunting. Especially when it is considered that in a system as complex as the DØ DAQ the same symptoms can mean different problems over time.

Finding a new problem isn't difficult. daqAI leaves behind enough logging information to make this easy. A key indication is if the *Unknown* category of downtime is quite large. Log file investigations and some time in the control room on shift will quickly point out the class of problems. Unfortunately, the symptoms for the problem are often duplicated during normal running and it can often take a few days of testing to get the rules just right. Once that is correctly implemented an automated fix can be added. It was often found that what looked like single problems were, in fact, two types and required different fixes. This process can take again several days to sort out correctly.

In the long term the inability of the script to respond to changing conditions can lead to problems. If the underlying problem is fixed, but daqAI's script is not changed it can introduce dead time into the system by issuing run control commands where they are not required.

Lastly, we were perhaps naïve thinking that the sociology of the experiment was not something we would have to deal with. Many detector groups were reluctant to have anything but a shifter control their detector. daqAI quickly gained acceptance as a tool to identify problems, but took a longer before people were comfortable with it sending direct commands and manipulating the system.

### 3.4. Future Directions

There are two possible improvements to daqAI based on current experiences: sensitivity to the time domain, and automatic run-condition classification.

daqAI is not sensitive to the sequence in which things happen without resorting to the timer-counters mentioned above. The symptoms for a problem often evolve over time, or the differentiating fact is what happens in the initial 10 seconds after the data rate begins to fall. Some of the monitor data has only course time resolution – more than 5 seconds – but much of it has 1 second resolution.

The second improvement addresses the most time consuming aspect of daqAI problem identification. There are many automatic classification schemes for physics variables based on various figures of merit. Something similar could be designed for a daqAI like system. One could imagine adding the ability to monitor shifter actions. When the shifter took an action that clearly changed the state of the system it would be recorded along with the current system state. Enough statistics and the system may be able to being a model.

Both of these approaches, though interesting, would require a good deal of effort. Their implementation schedule has not yet been decided.

## 4. CONCLUSIONS

The monitor system based on a caching monitor server has proved to be a simple, robust, and easy to use monitor system for the DØ DAQ and other parts of the DØ online system. The key to its adoption by the rest of DØ was the easy with which one could communicate. The protocol was designed to be as simple as possible and thus gained the acceptance that other monitor systems didn't as readily. We believe it is important to have as few monitor systems in an experiment as possible as the monitor system is only as powerful as the data it is serving.

Many clients and displays were discussed in this paper. In particular, the daqAI auto recovery program has proved to be a unique use of this monitor data. When first implemented it helped DØ gain over 10% data-taking uptime, and in that sense was very successful. It correctly identified and fixed the most vexing problems. It continues to function, though DØ's DAQ system is much more stable that before and thus makes use of daqAI auto-recovery feature less frequently.

The most important lesson learned by our group during the design and implementation of these projects was the key to having monitor information easily and quickly available. The design allowed us to quickly add new monitor items in even some of the busiest environments. Rich and prompt monitor data is a good start to better experiment uptimes.